\documentclass[12pt]{article}
\usepackage{amstex}
\usepackage{amssymb}

\newfont{\bit}{cmbxti10 scaled 1728}

\renewcommand{\thefootnote}{\fnsymbol{footnote}}
\begin{document}
\newpage
\pagestyle{empty}

\begin{center}
{\LARGE {Geodesics for impulsive gravitational waves\\
 and \\
the multiplication of distributions\\ 
}}

\vspace{3cm}

{\large Herbert BALASIN
\footnote[2]{e-mail: hbalasin @@ email.tuwien.ac.at}
\footnote[3]{supported by the APART-program of the 
Austrian Academy of Sciences}
}\\ 
{\em
 Institut f\"ur Theoretische Physik, Technische Universit\"at Wien\\
 Wiedner Hauptstra{\ss}e 8--10, A - 1040 Wien, AUSTRIA 
}\\[.8cm]
\end{center}
\vspace{2cm}
\begin{abstract}
We consider particle trajectories in the gravitational field of
an impulsive pp-wave. Due to the distributional character of 
the wave profile one inevitably encounters
an ambiguous point value $\theta(0)$. We show that this ambiguity 
may be resolved by imposing covariant constancy of the square of the
tangent. Our result is consistent with Colombeau's multiplication
of distributions.

\end{abstract}

\rightline{UWThPh -- 1996 -- 46}
\rightline{TUW 96 -- 12}
\rightline{June 1996}

\renewcommand{\thefootnote}{\arabic{footnote}}
\setcounter{footnote}{0}
\newpage  
\pagebreak
\pagenumbering{arabic}
\pagestyle{plain}
\renewcommand{\baselinestretch}{1}
\small\normalsize 
\section*{\Large\bit Introduction}

Impulsive pp-waves prominently arise as ultrarelativistic limits of 
black-hole geometries \cite{AS,BaNa3,BaNa4,LouSa}. Since their curvature is

completely concentrated on a null hyperplane they are necessarily
distributional in nature. Penrose \cite{Pen} gave an intrinsic 
description of these geometries in terms of his ``cut and paste''
approach. A different approach to derive the AS-geometry, which
represents the ultrarelativistic limit of Schwarzschild, was given by 't Hooft
and Dray \cite{tHo}. They made use of the limit of geodesics in
Schwarzschild to derive Penroses junction conditions. We will somehow
take the opposite point of view in this work and 
consider the trajectories of massive and massless particles in 
geometries generated by an impulsive gravitational wave in their own right. 
It turns out that the calculation inevitably requires the
definition of the point value $\theta(0)$ at the location of the pulse,
thus leaving the usual framework of distribution theory. 
>From a physical point of view this would induce a loss of predictability 
since the above point value is completely arbitrary. The ambiguity
is however resolved if one imposes the physically sensible requirement
that a freely falling (geodetic) observer experiences a constant flow
of eigentime, i.e. that the norm of the tangent vector remains constant.
Our result may be reinterpreted in the sense of Colombeau's theory
of multiplication of distributions if one replaces distributional equality by
weak equivalence of Coloumbeau functions. \par
In section one we calculate the geodesics for an arbitrary impulsive
wave profile. The ambiguity that arises from $\theta(0)$ is resolved
by taking the covariant constancy of the tangent vector into account.
Section two gives a brief account of the basic notions of Colombeau's
new generalized functions together with an extension to arbitrary 
$C^\infty$ manifolds. Finally, in section three we look at the calculations
of section one from the point of view of Colombeau theory thereby justifying
the previously obtained results. 

\section*{\Large\bit 1) Geodesics in (impulsive) pp-wave geometries }

A pp-wave is usually characterized as
being a vacuum geometry which admits a covariantly constant null-vector 
$p^a$ \cite{JEK}. However, since we do not want to restrict ourselves
to vacuum geometries, we require that
the Ricci-tensor is proportional to the tensor square of $p^a$.
>From these requirements one shows \cite{SiGo,AiBa2} that the metric 
may be written as
\begin{equation}\label{impwave}
  g_{ab} = \eta_{ab} + f p_a p_b \quad \mbox{together with}\quad 
  (p\partial)f=0,
\end{equation}
where $f$ denotes the so called wave profile and $\eta_{ab}$ is the
flat part of the decomposition. Using the covariant 
derivative $\partial_a$
associated with $\eta_{ab}$ one obtains the difference tensor
\begin{equation}
  C^a{}_{bc} = \frac{1}{2}\left( p^a \partial_bf p_c + 
    p^a \partial_c f p_b  - \partial^a f p_b p_c \right)
\end{equation}
Choosing a conjugate null vector $\bar{p}^a$ (i.e. $\bar{p}\cdot p=-1$) 
allows us to decompose
\begin{displaymath}
  \partial_af = - \bar{p}_a \underbrace{(p\partial)f}_{=0} 
  - p_a (\bar{p}\partial)f + \tilde{\partial}_a f,
\end{displaymath}
where the tilde refers to quantities which are projected in the
orthogonal two space of $p^a$ and $\bar{p}^a$.
This decomposition in turn gives rise to
\begin{equation}
  \label{diff}
  C^a{}_{bc} = \frac{1}{2}\left( p^a \tilde{\partial}_bf p_c + 
    p^a \tilde{\partial}_c f p_b  - \tilde{\partial}^a f p_b p_c 
    - p^a p_b p_c (\bar{p}\partial)f \right).
  \end{equation}
With respect to an affine parameterization the geodesic equation 
becomes

\begin{displaymath}
  \ddot{x}^a + C^a{}_{bc}\dot{x}^b  \dot{x}^c = 0.
\end{displaymath}
Decomposition with respect to the conjugate null directions yields
\begin{eqnarray}\label{geo}
  &&p\ddot{x} =0,\nonumber\\
  &&\bar{p}\ddot{x} + \frac{1}{2}\left( 
    -2(p\dot{x}) (\dot{\tilde{x}}\tilde{\partial})f + 
    (p\dot{x})^2(\bar{p}\partial)f 
    \right)=0,\nonumber\\  
  &&\ddot{\tilde{x}} - \frac{1}{2} (p\dot{x})^2 \tilde{\partial}f =0.
\end{eqnarray}
The first line of (\ref{geo}) allows us to choose $px$ as affine parameter
unless we are moving in a $px=const$ surface. This choice leaves us with two
equations
\begin{eqnarray}
  \label{ngeo}
  &&(\bar{p}x)'' + \left(-x'\tilde{\partial}f + 
    \frac{1}{2}(\bar{p}\partial)f \right) =0,\nonumber\\
  &&\tilde{x}'' - \frac{1}{2} \tilde{\partial} f = 0,
\end{eqnarray}
where the prime denotes differentiation with respect to $px$.
Up to now our derivation did not make use of the fact that
we are considering impulsive wave profiles, i.e. 
$f(px,\tilde{x}) = \tilde{f}(\tilde{x}) \delta(px)$.
By a slight abuse of notation we will drop the tilde on the 
$f(\tilde{x})$ in the following.
Physically the above form implies that the geodesics are straight lines
above and below the null hyperplane of the pulse
\begin{eqnarray}\label{geoans}
  \bar{p}x &=& a(px) + b + \theta(px)\left( a^+ (px) + b^+ \right),\nonumber\\ 
  \tilde{x} &=& \tilde{a}(px) + \tilde{b} + \theta(px)
  \left( \tilde{a}^+ (px) + \tilde{b}^+ \right).
\end{eqnarray}
Plugging (\ref{geoans}) into the second line of (\ref{ngeo}) gives
$$
\delta(px)\tilde{a}^+ + \delta'(px)\tilde{b}^+ = 
\frac{1}{2}\delta(px)\tilde{\partial} f,
$$
which requires 

$$\tilde{b}^+=0 \quad\mbox{and}\quad 
\tilde{a}^+=\frac{1}{2}\tilde{\partial} f(\tilde{b}),
$$
whereas the first line 
$$
a^+\delta(px) + b^+ \delta'(px) = \frac{1}{2}\tilde{x}'\tilde{\partial} f
\delta(px) + \frac{1}{2} \delta'(px) f(\tilde{b})
$$
leaves us with 
$$
a^+=\frac{1}{2}\left( \tilde{a} + \theta(0)\tilde{a}^+\right)
\tilde{\partial} f(\tilde{b})\quad\mbox{and}\quad
b^+=\frac{1}{2}f(\tilde{b}).
$$
So the geodesic becomes
\begin{eqnarray}
  \label{geores}
  \bar{p}x &=& a(px) + b + \theta(px)\left( \frac{1}{2}(
    \tilde{a} + \theta(0)\tilde{a}^+ )\tilde{\partial} f(\tilde{b})(px) +
   \frac{1}{2}f(\tilde{b})\right),\nonumber\\
   \tilde{x} &=& \tilde{a} (px) + \tilde{b} + \theta(px)\frac{1}{2}
   \tilde{\partial} f(\tilde{b}) (px).
\end{eqnarray}
At first glance it seems as if we were done, would it not be for the
ominous factor $\theta(0)$ which destroys the predictability of the
scattering process.
However the physical requirement of the covariantly constant 
norm of the tangent vector will save the day, i.~e.~
$$
(\dot{x}\nabla)(g_{ab}\dot{x}^a\dot{x}^b) = 
2g_{ab}(\dot{x}\nabla)\dot{x}^a\dot{x}^b = 0 \quad\Rightarrow\quad 
g_{ab}\dot{x}^a\dot{x}^b = const
$$ 
For a general pp-wave this gives
\begin{eqnarray}
  \label{covconst}
&&\dot{x}^a\dot{x}^b(\eta_{ab} + fp_a p_b) = 
-2(p\dot{x})(\bar{p}\dot{x}) + \dot{\tilde{x}}^2 + f (p\dot{x})^2 = 
const.\nonumber\\
&&\quad\Rightarrow\quad -2(\bar{p}x') + (\tilde{x}')^2 + f = const
\end{eqnarray}
Inserting (\ref{geores}) into (\ref{covconst}) we obtain immediately
\begin{equation}
  -2a + \tilde{a}^2 + \theta(px) 
  \frac{1}{4}(\tilde{\partial}f)^2(\tilde{b})\left(1-2\theta(0)\right) 
 = const,
\end{equation}
which is only constant if we choose $\theta(0)=1/2$, thereby
restoring predictability.

\section*{\Large\bit 2) Weak equality and Colombeau's product of 
distributions}

Although the result of the previous section is physically 
satisfactory  it suffers from mathematical deficiencies. This is
due to the fact that  we implicitly multiplied discontinuous functions
with (singular) distributions resulting in the ambiguous point value
  $\theta(0)$. Therefore this section is devoted to a brief summary of 
Colombeau's new generalized functions \cite{Col1,Col2,ArBi}, 
together with an extension to arbitrary manifolds, since they
 provide a natural framework for the multiplication
of distributions on a mathematically rigorous basis. From the 
physical point of view Colombeau objects may be regarded as 
(arbitrary) regularisations of singular distributions. More 
precisely\footnote{For a motivation of the growth condition see \cite{Col1}} 
one considers one-parameter families $(f_\epsilon)_{0<\epsilon<1}$ 
of $C^\infty$ functions with moderate growth in the parameter $\epsilon$,
namely
\begin{eqnarray}\label{Mod}
  && {\mathcal E}_M = \{ (f_\epsilon)\vert f_\epsilon \in 
    C^\infty({\mathbb R}^n)\quad \forall K \subset {\mathbb R}^n compact, 
    \forall \alpha\in {\mathbb N}^n\quad  \\
  &&\hspace*{1cm}\exists\, N\in {\mathbb N},\exists\> \eta > 0,\exists\> c>0 
    \quad s.t. \sup_{x\in K}\vert D^\alpha f_\epsilon(x)\vert \leq 
  \frac{c}{\epsilon^N}\quad\forall 0<\epsilon< \eta\}.\nonumber
\end{eqnarray}
Addition, multiplication and differentiation are defined as pointwise
operations, turning ${\mathcal E}_M$ into an algebra.
$C^\infty$ functions $f$ are naturally embedded into ${\mathcal E}_M$ as 
constant families, i.e. $f_\epsilon=f$, whereas continuous functions or 
elements of 
${\mathcal D}'$ require the use of a so called ``mollifier'' 
$\varphi$ for their embedding
\begin{equation}\label{moll}
 f_\epsilon(x) = \int\!\! d^n\!y \frac{1}{\epsilon^n}
 \varphi\fracwithdelims(){y-x}{\epsilon}f(y)\qquad \int 
d^n\!y\varphi(y) =1,
\end{equation}
together with additional conditions on the momenta of $\varphi$ 
\cite{Col1,Col2}. 
With regard to distributions the above convolution integral is formal. 
In order to identify the different embeddings of $C^\infty$ functions  
one takes the quotient of the algebra ${\mathcal E}_M$ with respect to
the ideal 
\begin{eqnarray}
  && {\mathcal N} = \{ (f_\epsilon)\vert f_\epsilon \in 
  C^\infty({\mathbb R}^n)\quad
  \forall K \subset {\mathbb R}^n compact, \forall \alpha\in {\mathbb N}^n,  
   \forall N\in {\mathbb N}\\
  &&\hspace*{1cm}\exists\> \eta > 0,\exists\> c>0,\quad 
  s.t. \sup_{x\in K}\vert D^\alpha f_\epsilon(x)\vert \leq 
  c\epsilon^N\quad\forall\> 0<\epsilon< \eta\}.\nonumber
\end{eqnarray}
thereby obtaining the Colombeau algebra $\mathcal G$.
Elements of $\mathcal G$ are therefore equivalence classes of 
one-parameter families of $C^\infty$ functions. In order to make contact
with distribution theory one considers a coarser equivalence relation which 
is usually called weak equality or association. It is intuitively 
motivated by the fact that the limit $\epsilon\to 0$ (if it exists) should 
reproduce the corresponding distributional object. More precisely, two 
Colombeau-objects $(f_\epsilon)$ and $(g_\epsilon)$ are associated if
\begin{equation}\label{assoc}
  \lim_{\epsilon\to 0} \int\! d^n\!x\left( f_\epsilon(x)-g_\epsilon(x) \right)
  \varphi(x)=0\qquad \forall \varphi\in C^\infty_0.
\end{equation}
Association behaves like equality on the level of distributions.
It is compatible with addition and differentiation and it allows
multiplication with $C^\infty$ functions. However, it does not respect
multiplication of Colombeau objects, as might have been expected.
The classical example in this context is provided by the powers of the 
$\theta$ function, which upon naive multiplication would lead to 
contradictions about the point value of 
$\theta(0)$. Specifically
$$
(\theta(x))^n = \theta(x) \Rightarrow n\theta(0)\theta'(x) = \theta'(x)
$$
which cannot hold for arbitrary $n$ since $\theta(0)$ is $n$-independent.
Looking at the above equality from the Colombeau point of view
we have
$$
(\theta(x))^n \approx \theta(x) \Rightarrow n (\theta(x))^{n-1}\theta'(x)
\approx \theta'(x)
$$
which holds separately for each $n$ and thereby allows us to replace 
$(\theta)^{n-1}\theta'$ by $(1/n)\theta'$. The contradiction is
avoided because multiplication by theta would break the association. 

The above concepts made explicit use of the additive (group) structure of 
${\mathbb R}^n$. Specifically the convolution integral used for the
embedding of continuous functions and distributions in general lacks
generalization to arbitrary manifolds. In the following we will 
give a coordinate--independent formulation of the Colombeau 
algebra\footnote{A more detailed version will be given in \cite{Bal2}}.
To begin with let us briefly recall the definition of distributions
on arbitrary (paracompact) $C^\infty$ manifolds.
The natural generalization of distribution space as
the dual of a suitable test-function space \cite{GeSh} is achieved by 
replacing 
test functions by test forms. That is we define a distribution as (continuous)
a linear functional on the space of $C^\infty$ n-forms with compact 
support together with the usual locally convex vector space topology
\cite{ChoBr}.
Locally integrable functions $f$ give rise to regular functionals via
$$
\tilde\varphi \mapsto (f,\tilde{\varphi}) := 
\int\limits_{M} f\tilde{\varphi} \qquad 
\forall\tilde{\varphi}\in C_0^\infty(M)
$$
The natural generalization of the derivative of a distribution
is given by the notion of the Lie-derivative with respect to
an arbitrary $C^\infty$ vector-field $X$, i.~e.~
$$
({\mathcal L}_X f,\tilde{\varphi}) := (-1) (f,{\mathcal L}_X\tilde{\varphi}).
$$
The above definition shows that distribution space does not require
additional structure but is purely a concept
which depends on the differentiable structure of the manifold.
It reduces to the usual notion of distribution space on ${\mathbb R}^n$
if one makes use of the natural volume form $d^n x$ and decomposes
every test-form $\tilde{\varphi}=\varphi(x)d^n\! x$.
Let us now try to do the same for the Colombeau algebra.
The $C^\infty$ functions of moderate growth in $\epsilon$ are easily 
generalized
\begin{eqnarray*}
{\mathcal E}_M(M) &=& \{ (f_\epsilon) \in C^\infty(M) | \forall\>
K \subset M\>\mbox{compact}, 
\forall\>\{X_1,\dots,X_p\}\>\\
&&\hspace*{0.5cm}X_i\in\Gamma(TM),[X_i,X_j ]=0,\exists N\in{\mathbb N},\exists 
\eta> 0, 
\exists c>0\> \\
&&\hspace*{0.5cm} s.t. \sup_{x\in K}|{\mathcal L}_{X_1}\dots{\mathcal L}_{X_p}
f_\epsilon(x)|
\leq \frac{c}{\epsilon^N}\quad 0<\epsilon<\eta\}
\end{eqnarray*}
In the same manner we may extend the ideal ${\mathcal N}$ 
\begin{eqnarray*}
{\mathcal N}(M) &=& \{ (f_\epsilon) \in C^\infty(M) | \forall\>
K \subset M\>\mbox{compact},\forall\>\{X_1,\dots,X_p\}\> \\
&&\hspace*{0.5cm}X_i\in\Gamma(TM),
 [X_i,X_j]=0,\forall q\in{\mathbb N},\exists \eta> 0, \exists c>0\>\\
&&\hspace*{0.5cm} s.t. \sup_{x\in K}|{\mathcal L}_{X_1}\dots
{\mathcal L}_{X_p}f_\epsilon(x)|
\leq c\epsilon^q\quad 0<\epsilon<\eta\},
\end{eqnarray*}
As usual the Colombeau-algebra ${\mathcal G}(M)$ is defined to be the 
quotient of ${\mathcal E}_M(M)$ with respect to ${\mathcal N}(M)$. 
In order to generalize the embedding of continuous functions 
and more generally of distributions, we have to find an analogue 
of the smoothing kernel $\varphi$. The immediate problem we are facing
is that the expression used in ${\mathbb R}^n$
\begin{equation}\label{falt}
f_\epsilon(x) = \int f(y)\varphi\fracwithdelims(){y-x}{\epsilon}
\frac{1}{\epsilon^n}d^n\! y
\end{equation}
makes explicit use of the linear structure of ${\mathbb R}^n$ as may
be seen from the argument of $\varphi$ in (\ref{falt}). Moreover, in
order to allow the identification of the two types of embeddings of
$C^\infty$ functions, as discussed at the beginning of this section, one 
has to require
$$
\int y^\alpha \varphi(y) d^n\! y = 0 \qquad 1\leq |\alpha | \leq q,
$$
for some finite but arbitrary $q\in {\mathbb N}$. Once again this expression
depends  explicitly on the chosen coordinates. At first glance it seems 
that the above conditions have to be modified \cite{ColMer} to allow a 
generalization to an
arbitrary manifold. However, if one makes use of the tangent bundle $TM$
of the manifold $M$ the above conditions may be taken as they are.
In order to show how this can be done let us consider local 
coordinates on $TM$ that are induced by coordinates on $M$,
i.~e.~ a bundle chart.
Let $(U,x)$ denote a local coordinate system of $M$ then $(\pi^{-1}(U),
(x,\lambda))$ will denote the corresponding bundle chart, where
$TM@>\pi>>M$ refers to the canonical projection and $\lambda$ to the 
coordinates along the fibres. Diffeomorphisms $M @>\mu>>M$ induce
fibre-preserving diffeomorphisms on 
$TM@>(\mu,(\partial\mu/\partial x))>>TM$, namely 
so called bundle morphisms. The smoothing kernels will now be taken to
be $C^\infty$ n-forms on $TM$, which allow a local ADM-like 
representation
$$
\tilde{\varphi} = \varphi(x,\lambda)(d\lambda^i+N^i{}_jdx^j)^n.
$$
Making use of the embedding map 
$$
i_x :T_x M@>>> TM\qquad \lambda \mapsto (x,\lambda)
$$
we require
$\tilde{\varphi}$ to obey
\begin{eqnarray}\label{smoothk}
  &&\int\limits_{T_xM}\!\!i_x^*\tilde{\varphi} = 
  \int\limits_{T_xM}\!\!\varphi(x,\lambda)d^n\lambda=1,\qquad
  i_x^*\tilde{\varphi} \in \Omega_0(T_xM),\nonumber\\
  &&\int\limits_{T_xM}\!\!\lambda^\alpha i_x^*\tilde{\varphi} = 
  \int\limits_{T_xM}\!\!\lambda^\alpha\varphi(x,\lambda)d^n\lambda=0
  \qquad\forall\>1\leq |\alpha|\leq q.
\end{eqnarray}
All of the conditions (\ref{smoothk}) are invariant under 
$M$-diffeomorphisms, since the latter act linearly on the fibres.
Moreover, rescaling the smoothing kernel is a well defined concept, 
since the transformation
$$
\phi_\epsilon: (x,\lambda) \mapsto (x,\frac{1}{\epsilon}\lambda)\qquad
\tilde{\varphi}_\epsilon := \phi_\epsilon^*\tilde{\varphi}
$$
is a specific case of the left action of the structure group 
$GL(n,{\mathbb R})$ of $TM$.
Having solved the first part of the problem we now have to decide
how to generalize the convolution integral necessary for the embedding 
of the continuous functions in the Colombeau algebra. Let us therefore 
choose a local coordinate system on $M$, which at the same time induces
coordinates on every tangent space in its domain. Let us, moreover, denote 
the representative of a $C^\infty(M)$ function with respect to the above 
coordinates by $f(x)$. Identification of the local coordinates
on $M$ with those of the fibre attached at $x$ allows us to ``lift''
$f$ to $T_xM$ by defining the value of the lift at $\lambda$ to
be $f(x+\lambda)$. The corresponding smoothened function $f_\epsilon$ 
on $M$ will then be defined to be
$$   
f_\epsilon(x):=\int\limits_{T_xM}\!\!f(x+\lambda)\varphi(x,
\frac{\lambda}{\epsilon})\frac{1}{\epsilon^n}d^n\!\lambda
=\int\limits_{T_xM}\!\!f(x+\epsilon\lambda)\varphi(x,\lambda)d^n\!\lambda
$$
This definition is not coordinate invariant because the action of an
$M$-diffeomorphism $\mu$ yields
\begin{eqnarray}\label{coinv1}
  \bar{f}_\epsilon(\bar{x}) &=& \int\limits_{T_{\bar{x}}M}\!\!\bar{f}
  (\bar{x}+\bar{\lambda})\bar{\varphi}(\bar{x},
  \frac{\bar{\lambda}}{\epsilon})\frac{1}{\epsilon^n}
  d^n\!\bar{\lambda}
  =\int\limits_{T_{\bar{x}}M}\!\!\bar{f}(\bar{x}+\epsilon\bar{\lambda})
  \bar{\varphi}(\bar{x},\bar{\lambda})d^n\!\bar{\lambda}\nonumber\\
  &=&\int\limits_{T_xM}\!\!f(\mu^{-1}(\mu(x)+\epsilon
  \frac{\partial\mu}{\partial x}\lambda)\varphi(x,\lambda)d^n\!\lambda,
\end{eqnarray}
which obviously differs from the definition in the unbarred system 
\begin{equation}\label{coinv2}
  f_\epsilon(x)=\int\limits_{T_xM}\!\!f(x+\epsilon\lambda)
  \varphi(x,\lambda)d^n\!\lambda
\end{equation}
The difference of (\ref{coinv1}) and (\ref{coinv2}) is of order
$\epsilon^{q+1}$ if all the momenta of $\varphi$ up to order $q$ vanish,
which is a necessary condition to belong to the ideal $\mathcal N$.
Thus the corresponding Colombeau object is well-defined since the 
coordinate change only generates a motion within its equivalence class. 
Finally, the concept of association is also without problems, i.~e.~
$$
(f_\epsilon) \approx (g_\epsilon)\>\mbox{iff}\> 
\lim_{\epsilon\to 0}\int\limits_M(f_\epsilon - g_\epsilon)\tilde{\varphi}=0
\>\forall \tilde{\varphi}\in\Omega_0(M).
$$ 
We will say that the Colombeau object $(f_\epsilon)$ gives rise to a
distribution $T$ if 
$$
(T,\tilde{\varphi}):=\lim_{\epsilon\to 0} \int\limits_M f_\epsilon 
\tilde{\varphi}
$$
defines a (continuous) linear functional over the space of test-forms.
The correspondence will in general be many to one. It is easy
to see that two different distributions with associated Colombeau
objects require the latter to be also different. 
As in the case of ${\mathbb R}^n$, association of Colombeau-objects
becomes equality on the level of distributions and all vector space
operations together with Lie-derivatives do not break the association.

\section*{\Large\bit 3) Geodesic equation and Colombeau theory}
Let us now look at the geodesic equation from the point of view 
of Colombeau theory. That is, replace the delta-function
appearing in (\ref{impwave}) by an arbitrary regularisation 
$\delta$ 
and do the same for the $\theta$'s that enter in (\ref{geoans}).
Taking into account the three Killing-vectors of the general 
impulsive wave-profile \cite{AiBa2,AiBa3}
\begin{equation}  \label{killings}
  \xi_1^a = p^a\quad \xi_{\tilde{t}}^a = (\tilde{t}x )p^a - 
  (px)\tilde{t}^a ,
\end{equation}
where $\tilde{t}$ denotes an arbitrary vector in the subspace transversal 
to $p^a,\bar{p}^a$, yields
\begin{eqnarray}\label{coconst}
  && p\dot{x} \approx const\nonumber\\
  && (\tilde{t}x )p\dot{x} - (px)(\tilde{t}\dot{x})\approx const.
\end{eqnarray}
The first line of (\ref{coconst}) just tells us that $px$ is an 
affine parameter, while
the second line fixes the transversal shift $\tilde{b}^+$ in (\ref{geoans}) 
to be zero.
The requirement that the length of tangent vector is covariantly
constant along the geodesic becomes
\begin{equation}
  -2a + \tilde{a}^2 + (-2a^+ + (\tilde{a}^+)^2)\theta +  
  ( -2b^+ + f(\tilde{b}) )\theta' \approx const,
\end{equation}
which implies that the coefficients of $\theta$ and $\theta'$
have to be zero 
\begin{eqnarray}
  a^+ &=& \tilde{a}\tilde{a}^+ + \frac{1}{2}(\tilde{a}^+)^2,\nonumber\\
  b^+ &=& \frac{1}{2} f(\tilde{b}).
\end{eqnarray}
>From the projection of the geodesic onto the orthogonal complement
of $p^a$ and $\bar{p}^a$
\begin{equation}
  \tilde{x}'' - \frac{1}{2}\tilde{\partial}f \delta \approx 0
\end{equation}
together with the above results we obtain
$$
\tilde{a}^+ = \frac{1}{2}\tilde{\partial}f(\tilde{b})
$$
Since all the plus-parameters are determined in terms of the
initial data, the remaining equation has to be satisfied consistently.
\begin{eqnarray}\label{consist}
  &&(\bar{p}x)'' -  (x'\tilde{\partial})f\>\delta - 
   \frac{1}{2} f\>\delta'\approx 0\nonumber\\
  &&(a^+(px)+ b^+)\theta'' + 2 a^+\theta' 
  - \frac{1}{2}(\tilde{a} + ((px)\theta)' (\tilde{a}^+
  \tilde{\partial})f)\>\delta - \frac{1}{2} f(\tilde{b})\>\delta' 
  \approx 0\nonumber\\
  &&b^+\theta'' +  a^+\theta' -\frac{1}{2} (\tilde{a}\tilde{\partial})
  f(\tilde{b})\theta' -\frac{1}{2}(\tilde{a}^+
  \tilde{\partial})f\>((px)\theta)'\delta - \frac{1}{2} f(\tilde{b})\theta''
  \approx 0\nonumber\\
  &&\frac{1}{8}(\tilde{\partial}f)^2(\tilde{b})\theta'
  -\frac{1}{4} (\tilde{\partial}f(\tilde{b})
  \tilde{\partial})f\>((px)\theta)'\delta \approx 0.
\end{eqnarray}
Taking into account that $((px)\theta)'$ is also a Theta-function,
which will be denoted by $\hat{\theta}$ one has 
$\hat{\theta}\delta \approx C \theta'$ as may be seen from 
\begin{eqnarray}\label{ambi}
  (\hat{\theta}\delta,\varphi) &=&\int\limits_{-\infty}^\infty\!du
  \frac{1}{\epsilon}\phi\fracwithdelims(){u}{\epsilon}
  \int\limits_{-\infty}^u\!dv\frac{1}{\epsilon}
  \psi\fracwithdelims(){v}{\epsilon}\varphi(u)
  \qquad\phi,\psi\in C^\infty_0\nonumber\\
&=&\int\limits_{-\infty}^\infty\!du\phi(u)\int\limits_{-\infty}^u\!dv\psi(v)
\varphi(\epsilon u) = C\varphi(0) + {\mathcal O}(\epsilon).
\end{eqnarray}
In (\ref{ambi}) $\phi$ and $\psi$ respectively
denote the regularisations of $\delta$ and $\hat{\theta}'$.
Therefore (\ref{consist}) becomes
$$
\theta' \frac{1}{8}(\tilde{\partial}f)^2(\tilde{b})(1-2C) \approx 0,
$$
which requires $C$ to be $1/2$. However, this is precisely the type of 
condition we encountered in section one. It restricts the regularisations
one is allowed to choose for $\delta$ and $\theta$ in order to obtain
a consistent (regularisation--independent) result.

\vfill
\noindent
{\bf Acknowledgement:} The author wants to thank Peter Aichelburg for 
many useful discussions and the critical reading of the manuscript, and 
Michael Oberguggenberger and Michael Kunzinger for introducing him
to Colombeau theory during his stay in Innsbruck in February 1996.  
\newpage
\section*{\Large\bit Conclusion}

In this work we solved the geodesic equation for arbitrary impulsive 
pp-wave geometries. Due to the singular character of the wave profile,
which contains a delta-function, one inevitably leaves the framework of
(classical) distribution theory. This fact manifests itself in the
appearance of the ambiguous pointvalue $\theta(0)$. However, since
we made use of an affine parametrization, the length of the tangent vector 
remains covariantly constant along the geodesic. This requirement fixed
$\theta(0)$ to be $1/2$. In order to justify our result mathematically 
we made use of the recently developed framework of Colombeau's new 
generalized functions, which is designed to allow a (consistent) 
multiplication of distributions. Although the original definition of the 
Colombeau algebra $\mathcal G$ made use of ${\mathbb R}^n$-specific concepts 
we showed that a generalisation to an arbitrary manifold $M$ is possible by 
employing the corresponding tangent bundle $TM$. Finally we showed that our 
condition on $\theta(0)$ may be justified from the point of view of
Colombeau-theory as condition on the ``regularizations'' used for the pulse 
and the geodetic trajectory, respectively. \par 
As a next step we will investigate semiclassical scattering,
namely various wave equations in a general impulsive pp-background with
specific emphasis on those geometries arising as ultrarelativistic 
limits of black-hole geometries.


\newpage
\begin{thebibliography}{99}
\bibitem{AS}Aichelburg P and Sexl R, 
{\em Gen.~Rel.~Grav.~}{\bf 2} (1971) 303.
\bibitem{BaNa3} Balasin H and Nachbagauer H, {\em Class.~Quantum Grav.~}
{\bf 12}, 707 (1995).
\bibitem{BaNa4} Balasin H and Nachbagauer H, {\em Class.~Quantum Grav.~}
{\bf 13}, 731 (1996).
\bibitem{LouSa} Loust\'o C and S\'anchez N 1992  
  {\em Nucl.~Phys.~}{\bf B383} 377.  
\bibitem{Pen}Penrose R, {\em General Relativity:  Papers in the Honour of 
J.~L.~Synge}, 101,Clarendon Press, Oxford  (1972).
\bibitem{tHo} 't Hooft G and Dray T, {\em Nucl.~Phys.~} {\bf B253}, 
173 (1985).
\bibitem{JEK} Jordan P Ehlers J and Kundt W, {\em Akad.~Wiss.~Lit. (Mainz)
Abhandl.~Math.-Nat.~Kl.~} {\bf 2}, 21 (1960). 
\bibitem{SiGo} Sippel R and Goenner H, {\em Gen.~Rel.~Grav.~ }
{\bf 18}, 1229 (1986).
\bibitem{AiBa2}Aichelburg P and Balasin H, {\em Class.~Quantum Grav.~}
{\bf 13}, 723 (1996).
\bibitem{Col1} Colombeau J, {\em New Generalized Functions and 
Multiplication of Distributions }
Mathematics Studies {\bf 84}, North Holland (1984).
\bibitem{Col2}Colombeau J, {\em Multiplication of Distributions }
{\bf LNM 1532}, Springer (1992).
\bibitem{ArBi}Aragona J and Biagioni H, {\em Analysis Mathematica}
{\bf 17}, 75 (1991). 
\bibitem{GeSh}Gel'fand I and Shilov G, {\em Generalized Functions}
{\bf Vol.~1}, Academic Press (1964).
\bibitem{ChoBr} Choquet-Bruhat Y, Morette-DeWitt C and Dillard-Bleick M,
{\em Analysis Manifolds and Physics}, North-Holland (1982).
\bibitem{Bal2}Balasin H, {\em Generalization of Colombeau's algebra to 
arbitrary $C^\infty$ manifolds}, in preparation.
\bibitem{ColMer}Colombeau J and Meril A, {\em Journal of Mathematical 
Analysis and Applications} {\bf 186}, 357-364 (1994).
\bibitem{AiBa3}Aichelburg P and Balasin H, {\em Generalized Symmetries
 of Impulsive Gravitational Waves }, gr-qc/9607045.
\end{thebibliography}
\end{document}